# Broadband Transistor-Injected Dual Doping Quantum Cascade Laser

**ZHIYUAN LIN,[1] ZHUORAN WANG,[1] GUOHUI YUAN,[1,*] AND JEAN-PIERRE LEBURTON[2]**

[1]*School of Information and Communication Engineering, University of Electronic Science and Technology of China, Chengdu, Sichuan, China, 610054*
[2]*Micro and Nanotechnology Lab, University of Illinois at Urbana-Champaign, Urbana, IL, USA, 61801*
*\*Corresponding author: yuanguohui@uestc.edu.cn*

**A novel design-friendly device called the transistor-injected dual doping quantum cascade laser (TI-D²QCL) with two different doping in each stack of a homogeneous superlattice is proposed. By adjusting the base-emitter bias $V_{be}$ of the bipolar transistor to supply electrons in the dual doping regions, charge quasi-neutrality can be achieved to generate different optical transitions in each cascading superlattice stack. These transitions are then stacked and amplified to contribute to a broad flat gain spectrum. Model calculations of a designed TI- D²QCL show that a broad flat gain spectrum ranging from 9.41um to 12.01um with a relative bandwidth of 0.24 can be obtained, indicating that the TI- D²QCL with dual doping pattern may open a new pathway to the appealing applications in both MIR and THz frequency ranges, from wideband optical generations to advanced frequency comb technologies.**

http://xxx

Interest in quantum cascade lasers (QCLs) based frequency comb generations [1] in both mid-infrared (MIR) [2, 3] and terahertz (THz) frequencies [4-6] has intensified significantly [1-9] in recent years, due to their crucial roles in advancing groundbreaking technologies ranging from high-precision metrology, spectroscopy to frequency synthesis [1, 10]. In order to achieve frequency comb generations with multiple modes over a wide wavelength range, QCLs with broad gain spectra, for which homogeneous cascade structures made of identical superlattice cells (SLCs) with bound-to-continuum[11], dual-to-continuum [12], and continuum-to-continuum [13] optical transitions have been proposed, are required in the first priority. However, the gain spectra of these devices exhibit generally Lorentzian line shapes because of the thermal-induced homogeneous broadening. Thus it is challenging for these designs to meet the second essential requirement of frequency comb operation of the QCLs -- the flat gain spectrum characteristic, which, without loss of generality, has been illustrated [1-3, 5, 14] to play a critical role in minimizing the gain induced dispersion over the lasing MIR or THz window, leading to the achievements of evenly mode spacing frequency comb operations in QCLs.

It is well known that by integrating multiple optically active cores containing various repetitions of individual SLCs into a heterogeneous superlattice (SL) structure [2, 5, 15, 16], sharp optical transitions with diverse gain peaks emerge into a flat broadband gain spectrum. Although wideband MIR [2, 15] and THz [5, 16] emissions have been demonstrated, design and fabrication of the heterogeneous QCL (HQCL) made of different SL stacks faces several challenges. The first one, of most significance, but also time consuming, is the implementation of band structure engineering to devise optically active cores with various lasing frequencies, at the same time as optimizing the carrier injection QCL regions with multiple minibands that provide efficient electron couplings between successive active SLCs in each SL stack. Between adjacent SL stacks, quantum state alignment needs also to be optimized to enhance electron coupling rate for improved QCL output performance. Meanwhile, in all SL stacks, the requirements of charge quasi-neutrality need to be fulfilled to provide maximum overlap between cells within the same stack. Otherwise space charge effect will result in a nonlinear variation of the electric potential that detunes the gain spectra from cell to cell [17] and prevents coherent optical emission. However, due to the strong coupling between the lasing energy and lasing intensity [17], all the above HQCL requirements can hardly be realized simultaneously, and additional etching and post-growth are sometimes necessary to obtain a stable operation [16], making the design and fabrication of the broad flat gain device challenging. Thus a more design-friendly MIR/THz emitter is urgently needed.

For this purpose, the newly proposed transistor-injected quantum cascade laser (TI-QCL) [17, 18] that decouples the lasing energy from the injection current intensity to the SL region, is ideally suited. In this letter, we introduce a novel concept, the transistor-injected dual doping quantum cascade laser (TI- D²QCL) aimed for the broad flat gain spectrum in MIR and THz frequency ranges. Unlike the HQCL design made of different stacks of SLCs, our device consists of structurally identical SLCs, but with alternate doping integrated into one single SL stack. In each stack, a dual doping method is applied for all SLCs to generate different electric fields under specific injection current intensity, thereby resulting in one single SL stack with multiple optical transitions. By further integrating this dual doping SL stack (D²SLS) into the collector region of the TI-QCL, and by adjusting the base-emitter bias $V_{be}$ to maintain the charge quasi-neutrality in each D²SLS, the optical spectrum of each D²SLS is amplified into a broadband flat gain emission.

In Fig. 1 (a), the schematic of the TI-QCL [17] is given, in which the SL region is embedded into the n-doped collector region of a three-terminal n-p-n transistor structure. To overcome the triangle potential resulting from the p-doped base and the n-doped SL regions, as shown in Fig. 1 (b), a quantum impedance matching area was designed and inserted between the base and the SL regions, leading to an enhanced injection current density into the SL region above threshold [17]. In Fig. 1 (b), one sees that the collector-base bias $V_{cb}$ ascertains the minibands distribution to decide the lasing wavelength of the SL, whereas the base-emitter bias $V_{be}$ controls the flow of electrons injected into the SL region, thereby determining the lasing intensity of the TI-QCL. By adjusting $V_{be}$ independently to maintain the charge quasi-neutrality in the SL region, maximum overlaps of gain spectra of all SLCs can be achieved for coherent lasing in MIR or THz window [17].

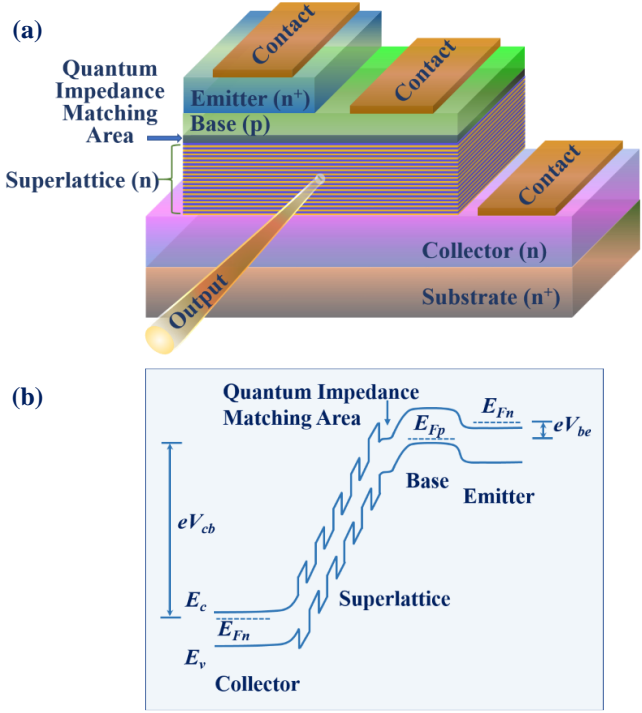

Figure 1. (a) Three-dimensional schematic of the TI-QCL structure. (b) Energy band diagram across the TI-QCL in normal operation status with the base-emitter and the collector-base junctions under forward and reverse biases, respectively, i.e., $V_{be}= V_b-V_e> 0$ and $V_{cb}= V_c-V_b> 0$. $E_{Fn}$ and $E_{Fp}$ are the quasi-Fermi levels in corresponding regions, $e$ is the element charge, and $E_c$ and $E_v$ are the conduction and valence band edge, respectively.

Based on the TI-QCL structure, a broadband TI- $D^2$QCL is proposed by further designing the SL region of the TI-QCL, as shown in Fig. 2 (a). One sees that the SL region of the TI-QCL consists of $m$ periods of $D^2$SLS that each contains $n$ periods of SLCs. All SLCs are identical in terms of the layer structure and doping layer positions, as shown in Fig. 2 (b). For the doping levels in one $D^2$SLS, the first and the rest $n-1$ periods SLCs are doped with $N_{d,1}$ and $N_{d,2}$, respectively, with $N_{d,1} > N_{d,2}$. In order to achieve cascading characteristics among all $D^2$SLSs, as suggested in Ref. [17], the charge quasi-neutrality in each $D^2$SLS should be fulfilled, reading as:

$$\frac{d^2V}{dx^2} = -\frac{\rho}{\epsilon} = -\frac{e}{\epsilon}\left(N_d^+ - \frac{J_e}{q\langle v_s\rangle}\right) = 0 \quad (1)$$

Where $V$ is the potential, $\rho$ is the charge density, $e$ is element charge, $\epsilon$ is the material permittivity, $J_e$ is the current density in the $D^2$SLS region, $\langle v_s\rangle$ is the equivalent velocity of electron. Defining $\rho_L$ as the net charge density over one $D^2$SLS, one get:

$$\left\langle\frac{\rho_L}{\epsilon}\right\rangle = e\int_{x_i}^{x_i+L}\frac{1}{\epsilon}\left(N_d(x) - \frac{J_e}{e\langle v_s\rangle}\right)dx = 0 \quad (2)$$

For the sake of simplicity in the derivation, $\epsilon$, $J_e$, and $\langle v_s\rangle$ are all assumed to be constant over the whole $D^2$SLS structure. By substituting the dual doping levels and lengths of the doping regions into Eq. (2), we get,

$$J_e = e\langle v_s\rangle\frac{N_{d,1}L_d+(n-1)N_{d,2}L_d}{L} \quad (3)$$

where $L_d$ and $L$ are the doping region length of one SLC and the length of one $D^2$SLS, respectively.

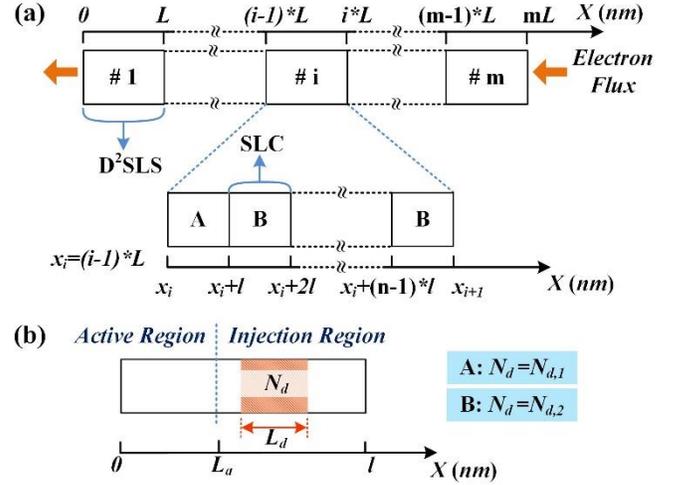

Figure 2. (a) Schematic of the $D^2$SLS structure of the TI- $D^2$QCL for broadband MIR and THz generation. The $l$ and $L$ designate the lengths of the SLC and $D^2$SLS respectively. The stripped shadow region with a length of $L_d$ is the doping region of each SLC, and the doping levels of A and B type SLC are $N_{d,1}$ and $N_{d,2}$, respectively.

It is easy to know that due to the negative electron flow and the positive ionized donors, the charge densities in the intrinsic and n-doped regions of all SLCs are negative and positive, respectively, leading to a periodic saw-tooth profile of the electric field $F$ in Fig. 3 (a). By integrating the charge density over the $j$th SLC, the electric fields difference $\Delta F$ between adjacent $j$th and $(j+1)$th $(1 < j \leq n)$ active regions in a $D^2$SLS can be derived as:

$$\Delta F = \frac{e}{\epsilon}\int_{x_i+(j-1)*l}^{x_i+j*l}\left[N_d(x) - \frac{J_e}{e\langle v_s\rangle}\right]dx$$
$$= \frac{e}{\epsilon}\left[\int_0^{L_d}N_{d,2}dx - \int_0^l\frac{J_e}{e\langle v_s\rangle}dx\right]$$
$$= -\frac{e(N_{d,1}-N_{d,2})L_d}{\epsilon n} \quad (4)$$

In Eq. 4, the integral limit $x_i + (j-1)*l$ is replaced by 0 due to the periodicity of the SLCs. From Eq. 4, one sees that from the 2nd to the $n$th SLC of the $i$th $D^2$SLS, and to the $1$st SLC of the $(i+1)$ $D^2$SLS, the $\Delta F$ is constant and SLC sequence independent, i.e., the electric fields in adjacent active regions experiences a linear decrease along these SLCs, as shown in Fig. 3 (a). In our previous paper [17], we showed that in TI-QCL, tunable optical transition energy $E_{OT}$ could be achieved when the SL region is biased under different electric fields, whereas charge quasi-neutrality is fulfilled. Assuming that the optical transition energy $E_{OT}$ as a linear function of the electric field $F$, it reads:

$$E_{OT} = \alpha F + \beta \quad (5)$$

where $\alpha$ and $\beta$ are fitting constant parameters for a specific SLC design. With an electric field variation of $\Delta F$, the $E_{OT}$ has a corresponding variation of $\Delta E_{OT} = \alpha\ \Delta F$. From Eq. (4), by simply designing the doping profile of $D^2$SLS, different $E_{OT}$ with a constant spatial energy of $\Delta E_{OT}$ can be obtained in each SLC, as shown in Fig. 3 (b). The varied $E_{OT}$ can then be merged into a broad flat MIR or THz gain spectrum by integrating $n$ periods of SLCs into one D²SLS. Finally, the wideband flat gain spectrum can be further amplified for coherent lasing by increasing the D²SLS number $m$.

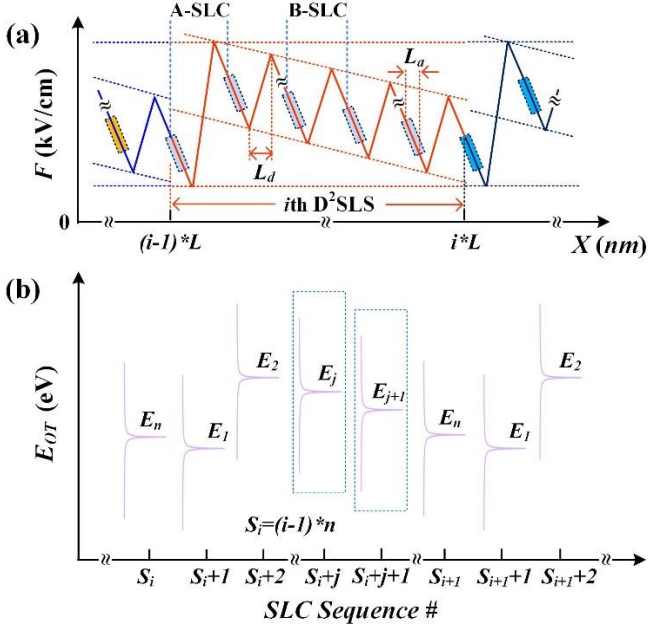

Figure 3. (a) Electric field over different SLCs in the $i$th D²SLS. The rectangles with dash outlines indicate the active regions of corresponding SLCs with same length of $L_a$, and $L_d$ is the doping region length of each SLC (b) Optical transition energy $E_{OT}$ spectra in successive SLCs.

In order to assess the performance of our proposed device, we simulate a TI- D²QCL with the following SLC layer sequences [48,**11**,54,**11**,19,**46**][30,**26**,30,**20**,28,**18**,30,**17**,34,**28**] (Å) [19], where the first and second sequences in the brackets are the active region and the injection region, respectively, and in which the bold and normal scripts denote the barriers Al$_{0.45}$Ga$_{0.55}$As and the wells GaAs, respectively, while the underscores depict the doped layers. By setting the parameters in Eq. (1) as follows: m=3, n=3, $N_{d,1} = 1.5 \times 10^{17} \, cm^{-3}$, $N_{d,2} = 1 \times 10^{17} \, cm^{-3}$, $\langle v_s \rangle = 1.25 \times 10^6 \, cm/s$, $\epsilon_{barrier} = 1.029 \times 10^{-12} \, Farad/cm$, and $\epsilon_{well} = 1.142 \times 10^{-12} \, Farad/cm$, the injection current $J_e$ that fulfills the charge quasi-neutrality in the D²SLS region and the $\Delta F$ are calculated to be 4.99 $kA/cm^2$ and 2.35 $kV/cm$, respectively. The $J_e$ is close to the value of the enhanced injection current (4.71 $kA/cm^2$) of the TI-QCL with quantum impedance match area [17], and comparable to the threshold current density (4 kA/cm2) in the conventional QCL with the original design in Ref. [19], indicating that the required $J_e$ could be achieved for lasing by independently tuning the base-emitter bias $V_{be}$ of the TI-D²QCL as suggested in Ref. [17]. By adjusting the base-emitter bias $V_{cb}$ in the TI- D²QCL, an additional electric field of 46.3 kV/cm is superimposed to the D²SLS region to provide efficient carrier cascading as in Refs. [17, 19]. Two 30 nm GaAs regions are added to both ends of the D²SLS to play the roles of injection and leaking regions [20], respectively. After obtaining the electric potential through Poisson equation, a Numerov Schrodinger solver with complex boundaries potential algorithm [20, 21] is used to calculate the quasi-bound states of the D²SLS structure, whereas the relevant 8 energy levels with the longest lifetimes in each SLC are selected as quasi-bound states in the SL region, as shown in Fig. 4. In this context, as in Ref. [17], the upper and lower lasing states are ascertained to be level 8 and levels 3 and 4, respectively, by analyzing the transition dipole moments of levels between upper and lower minibands.

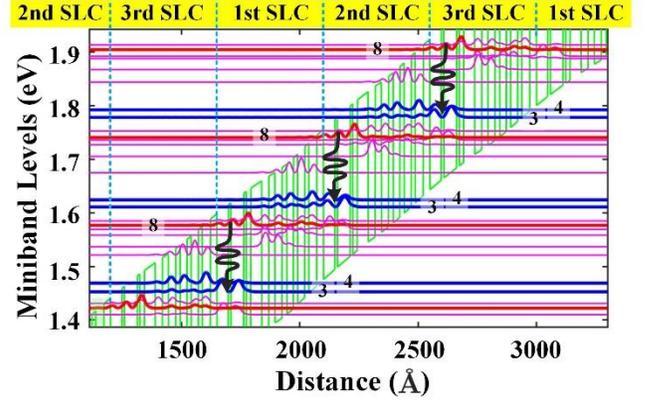

Figure 4. Miniband levels in the given D²SLS. The upper level 8 and lower lasing levels 3 and 4 are plotted in red and blue, respectively, the wavy dark arrows represent the optical transition processes.

To obtain the gain spectrum qualitatively, we consider gain broadening and calculate corresponding spectral gain cross section (SGCS) of the $j$th SLC by the following equation [5]:

$$g_{j,if} = \frac{2\pi e^2 d_{j,if}^2}{\epsilon_0 n_r l \lambda_{j,if}} \frac{\gamma}{\left(E_{j,if} - \frac{ch}{\lambda}\right)^2 + \gamma^2} \quad (6)$$

where $e, \epsilon_0, c$ and $h$ are the elementary charge, vacuum permittivity, speed of light and Planck constant, respectively, $d_{j,if}^2 = |\langle f|x|i\rangle|^2$ is the transition dipole moment square of upper lasing state $i$ and lower lasing state $f$ of the $j$th SLC that contribute to the total gain, $n_r$=3.27 is the effective refractive index, $l$=45 nm is the length of each SLC, $\lambda_{if}$ and $E_{if}$ are the wavelength and energy of corresponding optical transition, respectively, and $\gamma$ =6 meV is the gain broadening. By weighing the overlap factor $\frac{1}{n}$ for all $g_{j,if}$, the total SGCS is calculated as $g_t = \frac{1}{n}\sum_{j=1}^{3}\sum_{f=3}^{4} g_{j,if}$ and given in Fig. 5.

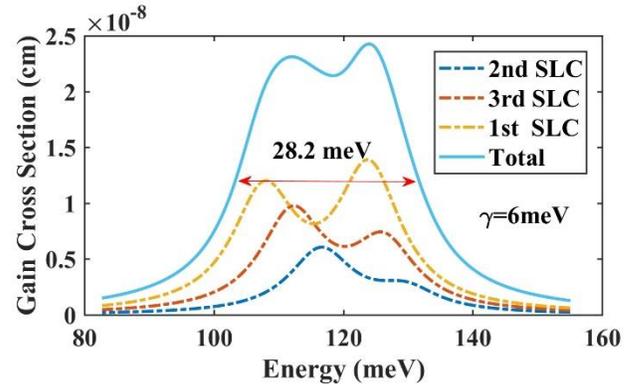

Figure 5. Total spectral gain cross section of the given D²SLS with the gain broadening $\gamma$=6 meV.

In Fig. 5, for each SLC, two SGCS peaks are observed because of the dual lower lasing levels (level 3 & 4) behavior as shown in Fig. 4. In one D²SLS, as predicted in Fig. 3 (b) and (c), relevant SGCS peaks have redshifts as the SLC moves from Cell 2 to Cell 3, and to Cell 1, in good agreement with our previous calculated results [17]. By taking into account all contributing SLCs, the total SGCS has a flat top and a full width at half-maximum (FWHM) of 28.2 meV (2.6 $\mu m$), from 103.4 meV (12.01$\mu m$) to 131.6 meV (9.41 $\mu m$), leading to a relative bandwidth of $\frac{\Delta E}{E} = \frac{28.2}{\frac{131.6+103.4}{2}} = 0.24$, which is comparable to the ones obtained with dual upper state to multiple lower states transition design [12] and HQCL [15].

Compared to HQCL designs, achieving broad flat spectrum through TI- D²QCL does not require involvements with complex band structure engineering nor space charge induced nonlinear potential in the SL region. In the meantime, in the TI- D²QCL, ultra-broad flat gain spectrum with even larger relative bandwidth could be achieved by further increasing the SLC number $n$ and carefully designing the $N_{d,1}$ and $N_{d,2}$ doping levels in one D²SLS, which can be further amplified for flat wideband coherent lasing by increasing the D²SLS number $m$. It is worth mentioning that this D²SLS approach can be also utilized in conventional unipolar QCL for achieving wideband flat spectrum as long as the charge quasi-neutrality can be maintained in the SL region.

To summarize, we proposed a TI-QCL with dual-doping structure as a design-friendly approach for broad flat gain spectrum generation. Our calculations showed that dual doping profile within structurally homogeneous SL consisted of single functional SLC design for the operation in the MIR window, resulted in a broad flat gain spectrum ranging from 9.41um to 12.01um with a relative bandwidth of 0.24 in a TI- D²QCL in which each D²SLS contains three SLCs. In addition, wideband THz generation can also be achieved by applying the dual doping approach to a demonstrated functional THz lasing SLC structure [4, 5, 16] with the D²SLS through the same roadmap. One can expect that by increasing the number $n$ and $m$ of SLC and D²SLS, respectively, and by optimizing the $N_{d,1}$ and $N_{d,2}$ doping levels, the TI- D²QCL could be a powerful approach for ultra-broad flat gain spectrum generations as well as in advanced MIR and THz frequency comb technologies.

**Funding.** National Natural Science Foundation of China (61575038, 61775030, 61571096); China Scholarship Council (2015-3022); and Fundamental Research Funds for the Central Universities (ZYGX2015J052).

**Disclosures.** The authors declare no conflicts of interest.